\documentstyle[12pt]{article}  

\advance\hoffset by -7mm  
\makeatletter
\@addtoreset{equation}{section}
\makeatother

\renewcommand{\title}[1]{\null\vspace{25mm}

\noindent{\Large{\bf #1}}\vspace{10mm}

\noindent {\large By }}
\newcommand{\authors}[1]{\noindent{\large #1}\vspace{3mm}

}
\newcommand{\address}[1]{\noindent #1\vspace{5mm}

}
\renewcommand{\abstract}[1]{\vspace{19mm}

\noindent{\small{\em Abstract.} #1}\vspace{2mm}

} 

\begin{document}
\begin{flushright}
RCG 96/09\\
SUSSEX-AST 96/7-6\\
astro-ph/9608042
\end{flushright}
\vspace*{-60pt}
\title{Density Perturbations from Two-field Inflation\footnote{Talk
presented at Les Journ\'ees Relativistes, Ascona, Switzerland,
26th-30th May 1996.}}
\authors{David Wands}
\address{School of Mathematical Studies, University of Portsmouth,\\
        Mercantile House, Hampshire Terrace, Portsmouth, PO1 2EG, United
        Kingdom}
\authors{and Juan Garc\'\i a-Bellido}
\address{Astronomy Centre, University of Sussex, Brighton, BN1 9QH,
        United Kingdom}
\abstract{ 
We discuss metric perturbations produced during a period of inflation
in the early universe where two scalar fields evolve.
The final scalar perturbation spectrum can be calculated in terms of the
perturbed expansion along neighbouring trajectories in field-space. In
the usual single field case this is fixed by the values of the fields
at horizon-crossing, but in the presence of more than one field there
is no longer a unique slow-roll trajectory. The presence of entropy as
well as adiabatic fluctuations means that the super-horizon-sized
metric perturbation $\zeta$ may no longer be conserved and the evolution
must be integrated along the whole of the subsequent trajectory.  In
general there is an inequality between the ratio of tensor to scalar
perturbations and the tilt of the gravitational wave spectrum, which
becomes an equality when only adiabatic perturbations are possible and
$\zeta$ is conserved.
}

\setcounter{section}{1}

Inflation was originally proposed as a mechanism to
produce a flat, isotropic and homogeneous universe. However it was
soon realised that vacuum fluctuations in a scalar field driving
inflation would also lead to the production of an approximately
scale-invariant spectrum of inhomogeneities. A good deal of progress
has been made in understanding and parameterising generic features of
the perturbation spectrum independently of the specific interaction
potential as long as it is driven by a single scalar
field~\cite{LL93}. Our intention here is to emphasise which features are
restricted to models of a single field evolving during inflation and
what happens when more than one field is present~\cite{SS96,GBW96}.

In single field inflation there is a single attractor trajectory,
the non-decaying mode for the homogeneous solution for the inflaton
field, $\phi(t)$. This has two important consequences. Firstly,
we can associate any particular time during inflation with a
particular value of the scalar field. Secondly, the
non-decaying mode of any perturbation on large scales (to be defined
below) corresponds to a perturbation in time
along this homogeneous solution, $\delta t\equiv\delta\phi/\dot\phi$.

We can always decompose the full time- and spatially-dependent
inflaton field into Fourier modes with comoving wavenumbers $k$, which
evolve independently in the linear approximation. Because the comoving
Hubble length, $H^{-1}/a$, shrinks during inflation any
Fourier mode will eventually be stretched far beyond the horizon, even
though it may start far within the Hubble scale. Assuming each mode
starts in the Bunch-Davies vacuum state on small scales ($k\gg aH$),
they will have an amplitude $\delta\phi\simeq H/2\pi$ at horizon
crossing\footnote{Strictly speaking, this is the perturbation in the
non-decaying asymptotic solution, evaluated when $aH = k$ for a
quasi-massless field ($m^2\ll H^2$).}, on spatially flat
hypersurfaces. If one makes a gauge transformation to a comoving
hypersurface (one with uniform $\phi$) we find the intrinsic curvature
perturbation is $^{(3)}\delta R \equiv 4{\cal R}\,k^2/a^2$, where the
metric perturbation, ${\cal R}$, is a particularly useful quantity. It
approaches a constant value on large scales~\cite{MFB},
\begin{equation}
\label{zeta1}
{\cal R} \to \zeta
 = \left( {H\delta\phi \over \dot\phi} \right)
 \qquad {\rm as} \qquad {k\over aH}\to0 \, ,
\end{equation}
which is simply the perturbation in the number of expansion times,
$\zeta\equiv\delta N=H\delta t$, where $\delta
t\equiv\delta\rho/\dot\rho$. It is important to emphasise that the
constancy of $\zeta$ on super-horizon scales is not dependent on the
slow-roll approximation, but is simply a consequence of the
perturbations being adiabatic. Indeed we can write down an expression
for the time derivative of $\zeta$~\cite{GBW96}
\begin{equation}
\dot\zeta
 = 3H \left( {\dot{p}\over\dot\rho}
             - {\delta p \over \delta\rho} \right) \zeta \, ,
\end{equation}
in the limit $k/aH\to0$, for a general fluid with pressure $p$ and
density $\rho$. Adiabatic perturbations can be represented by a time
perturbation $\delta t=\delta{p}/\dot{p}=\delta\rho/\dot\rho$.  
So long as there is a single fluid with any equation of state,
$p(\rho)$, the perturbations will be adiabatic and thus $\zeta$ is
conserved after inflation during the radiation
or matter dominated eras until $aH$ has decreased back to the value $k$ and
the mode re-enters the horizon~\cite{Lyth85}. The density contrast at
re-entry is given by $(\delta\rho/\rho)_{k=aH}=(2/5) \zeta$ during the
matter dominated era~\cite{LL93}. 
Even though the physics of re-heating the universe at the end of
inflation may be exceedingly complicated, $\zeta$ is conserved on
super-horizon scales when changes in the equation of state occur at a
fixed energy density, since $\zeta$ gives the intrinsic curvature
perturbation on the boundary~\cite{DM95}.

Each mode $k$ has a single horizon-crossing time during inflation when
$aH=k$ and the field $\phi$ has a value $\phi_*$. If we can determine
$\zeta(k)$ from observations of the microwave background sky or
large-scale structure, one can set about reconstructing the inflaton
potential $V(\phi)$ when these scales left the horizon, independently
not only of the details of re-heating, but also regardless of the
latter stages of inflation which can only affect smaller
scales.

It is straightforward to generalise Eq.~(\ref{zeta1}) to the case of
two scalar fields in the slow-roll limit,
\begin{equation}
\label{zeta2}
\zeta \simeq H\, \left( {\dot\phi\delta\phi + \dot\sigma\delta\sigma
                     \over \dot\phi^2 + \dot\sigma^2} \right)
 \qquad {\rm as} \qquad {k\over aH}\to0 \, ,
\end{equation}
where $\zeta$ now represents the metric perturbation on uniform
density hypersurfaces. However this quantity is not constant on
super-horizon scales due to the presence of entropy perturbations. There
is no longer a locally defined time perturbation since
$\delta\phi/\dot\phi\neq\delta\sigma/\dot\sigma$ in general, and we
have
\begin{equation}
\dot\zeta = {H\over2}
 \left({\delta\phi\over\dot\phi}-{\delta\sigma\over\dot\sigma}\right)
 {d\over dt} 
 \left({\dot\phi^2-\dot\sigma^2\over\dot\phi^2+\dot\sigma^2}\right)
\, .
\end{equation}
To know the value of $\zeta$ at the end of, or after, inflation
we must integrate $\zeta$ along the whole subsequent trajectory and
establish the final perturbation in the expansion time.  This is not
determined simply by the perturbation at horizon crossing because
different regions of the universe could subsequently follow radically
different trajectories.  With two fields present there may be many
different possible trajectories during inflation.  Even if we restrict
our analysis to the slow roll approximation, which reduces the
four-dimensional phase-space to a
two-dimensional field-space, the end of inflation surface, for
instance, now becomes a line rather than a single point in field
space.

On the other hand, if one can construct the function
$N(\phi,\sigma)=\int_\gamma H(\phi,\sigma)dt$ giving the total
expansion from any point to the end of inflation along each classical
trajectory $\gamma$, one can still in principle give
$\zeta_e\equiv\delta N$ in terms of the perturbation at horizon
crossing~\cite{SS96}. This is of course equivalent to integrating the
perturbation along the subsequent trajectory but gives an intuitively
simpler picture and allows one to work with only the homogeneous field
equations.  In previous work we have considered the perturbation
spectra generated during inflation in models with a separable interaction
between two scalar fields~\cite{GBW96}, a particular case of which is
scalar-tensor gravity where a dilaton (or Brans-Dicke) field is
non-minimally coupled to the metric~\cite{GBW95}.

If the second scalar field is held fixed ($\dot\sigma=0$) in a local
minimum of the potential then equation~(\ref{zeta2}) reduces to
equation~(\ref{zeta1}). In practice there is a single trajectory and
we can easily construct $N(\phi)$. This occurs in most models of
extended~\cite{extended} or hybrid~\cite{hybrid} inflation where
$\sigma$ has a large positive effective mass until some critical point
$\phi_c$, where $\sigma$ undergoes a phase transition and inflation
ends. However this relies on inflation ending rapidly at a first- or
second-order phase transition. Recently~\cite{Randall} it has been
argued that in a more ``natural'' model of hybrid inflation the two
fields should have similar bare masses $|m^2|\sim H^2\sim 1{\rm
TeV}$. If so, the phase transition may not complete rapidly and scales
crossing outside the horizon near to the phase transition may
correspond to large scales today. There is certainly more than one
possible trajectory close to $\phi_c$ where $\sigma$ is effectively
massless, and perturbations need not be purely adiabatic.  Fortunately
it is possible to give an approximate form for $N$ after the critical
point~\cite{GBLW} and we can also show that all the trajectories
converge on a single trajectory before the end of inflation. Thus by
the end of inflation we are left with adiabatic perturbations along
this trajectory. This is important as otherwise the final spectrum
could depend upon the detailed dynamics of re-heating, as mentioned
above.

In addition to the scalar perturbations, there is always a spectrum of
tensor perturbations (gravitational waves) generated during
inflation~\cite{GW}, whose amplitude is proportional to $H$ at horizon
crossing. The tilt of this spectrum is $n_T=2\dot{H}/H^2$ irrespective
of the matter content. Detection of such a spectrum of gravitational
waves with $-2<n_T<0$ would be a definitive test of inflation. In the
case of single field inflation, $n_T$ is related to the ratio of
tensor to scalar perturbations of the microwave background on large
angular scales, $R\simeq 6|n_T|$~\cite{LL93}. This has been proposed
as a test of inflation, but in fact this is a characteristic only of
purely adiabatic perturbations generated during inflation where
$\zeta$ is constant on super-horizon scales. In the presence of
entropy perturbations during inflation the equality becomes an
inequality $R< 6|n_T|$~\cite{Double,SS96,GBW96}.

\end{document}